\documentstyle[12pt]{article}
\def\ftoday{{\sl  \number\day \space\ifcase\month
\or Janvier\or F\'evrier\or Mars\or avril\or Mai
\or Juin\or Juillet\or Ao\^ut\or Septembre\or Octobre
\or Novembre \or D\'ecembre\fi
\space  \number\year}}

\makeatletter
\@addtoreset{equation}{section}
\makeatother

\def\non {\nonumber}
\def\6 {\partial}
\def\deb {\bar{\partial}}
\def\in {\int dzd\bar{z}}
\def\din{\displaystyle{\int dzd\bar{z}}}
\def\zb {\bar{z}}
\let\f=\frac
\def\fr {\frac{1}{1-\mu\bar{\mu}}}
\def\dfr{\displaystyle{\frac{1}{1-\mu\bar{\mu}}}}
\def\de {(\partial -\bar{\mu}\bar{\partial})}
\def\db {(\bar{\partial}-\mu\partial)}
\def\mb {\bar{\mu}}
\let\a=\alpha
\let\l=\lambda
\let\e=\epsilon
\let\t=\theta
\let\S=\Sigma
\let\d=\delta
\let\D=\Delta
\def\Wb {\bar{W}}
\def\Db {\bar{\Delta}}
\def\ds {\delta\Sigma}
\def\ab {\bar{\alpha}}
\def\lb {\bar{\lambda}}
\def\eb {\bar{\epsilon}}

\def\bb {\bar{b}}
\def\balf{b^{\a}}
\def\bab {\bar{b}^{\a}}
\def\2 {\frac{1}{2}}
\def\1 {(1-\mu\bar{\mu})}
\def\c {(c-\mu\bar{c})}
\def\cb {(\bar{c}-\bar{\mu}c)}

\def\X {J^X}
\def\L {J^{\l}}
\def\Lb {\bar{J}^{\l}}
\def\F {J^F}
\def\C {J^c}
\def\Cb {\bar{J}^{c}}
\def\E {J^{\e}}
\def\Eb {\bar{J}^{\e}}
\def\B {{\cal B}}

\renewcommand{\AA}{{\cal A}}

\def\Lp{\displaystyle{\biggl(}}
\def\Rp{\displaystyle{\biggr)}}

\newcommand{\complex}{{\kern .1em {\raise .47ex
\hbox {$\scriptscriptstyle |$}}
    \kern -.4em {\rm C}}}
\newcommand{\real}{{{\rm I} \kern -.19em {\rm R}}}
\newcommand{\rational}{{\kern .1em {\raise .47ex
\hbox{$\scripscriptstyle |$}}
    \kern -.35em {\rm Q}}}
\renewcommand{\natural}{{\vrule height 1.6ex width
.05em depth 0ex \kern -.35em {\rm N}}}

\newcommand{\sla}{\raise.15ex\hbox{$/$}\kern -.57em}
\newcommand{\twiddle}{\lower.9ex\rlap{$\kern -.1em\scriptstyle\sim$}}

\newcommand{\eq}{\begin{equation}}
\newcommand{\eqn}[1]{\label{#1}\end{equation}}
\newcommand{\eea}{\end{eqnarray}}
\newcommand{\eqa}{\begin{eqnarray}}
\newcommand{\eqan}[1]{\label{#1}\end{eqnarray}}
\newcommand{\ba}{\begin{array}}
\newcommand{\ea}{\end{array}}
\newcommand{\eqac}{\begin{equation}\begin{array}{rcl}}
\newcommand{\eqacn}[1]{\end{array}\label{#1}\end{equation}}

\def\beq {\begin{equation}}
\def\eeq {\end{equation}}
\def\beqa {\begin{eqnarray}}
\def\eeqa {\end{eqnarray}}
\def\non {\nonumber}

\begin{document}
\begin{titlepage}
\topskip1cm
\flushright{hep-th/9312087\\
CERN-TH-7110/93\\
TUW-93-27}\\[3cm]
\begin{center}{\huge\bf
Vector supersymmetry of the superstring in the super-Beltrami parametrization
\footnote{Supported in part by the "Fonds zur F\"orderung der
Wissenschaftlichen Forschung", M008-Lise Meitner Fellowship}
\\[2cm]}{
\Large A. Boresch$^{\#}$, M. Schweda$^*$ \\and S.P. Sorella$^*$
\\[1cm]}
$^{\#}$CERN, Theory Division\\
CH-1211, Geneva 23 (Switzerland)\\[0.5cm]
$^*$Institut f\"ur Theoretische Physik\\
Technische Universit\"at Wien\\
Wiedner Hauptstra{\ss}e 8-10\\
1040 Wien (Austria)\\[2cm]
\end{center}
\begin{abstract}

The vector supersymmetry recently found for the bosonic string is
generalized to superstring theories quantized in the super-Beltrami
parametrization.
\end{abstract}
\flushleft{CERN-TH-7110/93\\
TUW-93-27\\
December 1993}
\end{titlepage}

\section{Introduction}
Topological models \cite{witt,schw,birm} are known to be characterized
by a supersymmetric algebra of the Wess-Zumino type
whose generators are identified respectively with the BRS symmetry and
with the vector supersymmetry carrying a Lorentz index \cite{gier}.

This supersymmetric structure turns out to be very useful in
order to discuss the perturbative renormalization of these models.
Indeed, as shown in \cite{guad}, it
provides a simple way for solving the descent equations corresponding
to the integrated BRS cohomology, yielding then an algebraic
characterization of anomalies and invariant counterterms for both
Schwarz \cite{lucc} and Witten's type \cite{rako} topological models.

This paper is the continuation of a recent work \cite{sore} where
the vector supersymmetry has been established also in the case
of the bosonic string quantized in the Beltrami parametrization
\cite{baul,becc}.
In particular we show that the vector supersymmetry can be
generalized to the $(1,1)$ and $(1,0)$ superstring theories
and that, as in the bosonic case, it can be used for
an algebraic characterization of the superdiffeomorphism anomaly.

In the following we shall adopt the so called super-Beltrami parametrization.
We will make use of the results of \cite{bbg,deld}
where this parametrization has been developed in
a superfield as well as in a component field formalism.
Let us recall that the Beltrami parametrization,
introduced by \cite{baul}, has been shown \cite{becc}
to be the most natural parametrization which exhibits the
holomorphic factorization of the Green functions according to the
Belavin-Polyakov-Zamolodchikov scheme \cite{bela}. This holds true also for
the super-Beltrami parametrization \cite{deld}.

The work is organized as follows. Sect.2 is devoted to a brief summary of
the BRS quantization of the superstring in the super-Beltrami
parametrization. In Sect.3 we introduce the
vector supersymmetry and we study the related Ward identities.
Finally, in Sect.4 we solve the descent equations for the
superdiffeomorphism anomaly.
All the calculations are done for the case of the $(1,1)$
superstring, the $(1,0)$ case being easily recovered by means of
a simple truncation procedure, as explained in Sect.2.

\section{BRS-quantization of the superstring}
Following \cite{deld}, the superstring action in component fields and in
the Wess-Zumino (WZ) gauge is given by:
\eq\ba{rl}
 - 2iS_{inv}=  &\!\! \din  \dfr \Lp \de X \db X \Rp  \\
 +  &\!\! \din  \dfr \Lp \a\l\de X + \ab\lb\db X +\2 \a\l\ab\lb \Rp \\
 +  &\!\! \din  \Lp \l\db\l +\lb\de\lb +\1 F^2  \Rp \ .
\ea\eqn{invaction}
with
\beq
 \partial = \partial_{z} \ , \qquad
 {\bar \partial} = \partial_{\bar z} \ .
\eeq
Here the $\{ X \}$ are the string coordinates, the $\{ \l \}$ denote
the fermionic superpartners and the $\{ F \}$ are auxiliary fields.
$\mu$ stands for
the conventional (i.e. bosonic) Beltrami differential and $\a$, also
called "Beltramino", is its fermionic superpartner. The doublets $(\mu ,\a)$
and $(\mb ,\ab)$ characterize the super-Beltrami parametrization which is a
parametrization of the metric in the $2d$-superspace \cite{deld}.
For the two dimensional world sheet the line element takes the form
\beq
ds^2 \propto |dz+\mu d\zb |^2 \ .
\eeq
The quadratic part in the fields $\{ X \}$ of the expression
(\ref{invaction})
identifies the bosonic string whereas the quadratic terms in
$\{ \l \}$ are the corresponding fermionic counterpart.
Looking at the $(\l X)$-terms one can see that the
Beltramino $\a$ plays the role of the gravitino or Rarita-Schwinger field
which is present in other formulations of the superstring action.

Expression (\ref{invaction}) represents the full $(1,1)$ superstring theory.
According to \cite{deld}, the corresponding $(1,0)$ version
is easily obtaind by means of the truncation $(\ab =\lb =F=0)$, meaning
that the supersymmetry is present only in one sector while in
the other one has just the bosonic string.
In the following we will always refer to the $(1,1)$ model, all the
results being easily extended to the $(1,0)$ case by using the
above truncation procedure.

The classical action (\ref{invaction}) turns out to be invariant under the
superdiffeomorphism transformations which are expressed by the following
BRS transformations \cite{deld}
\beqa
sX&=&\fr [\c\6 X+\cb \deb X+\2 \a\cb\l ]+\2 \e\l \ ,\\
s\l&=&\fr [\c\6 \l +\cb \deb\l]
        +\2 \left[\6 c-\frac{\cb}{\1 }\6 \mu\right]\l \non\\
  & &\qquad\qquad +\2 \left[\frac{\a\cb}{\1 }+\e\right] D_z X-
\frac{i}{2}\left[\frac{\ab\c}{\1 }+\eb\right] F \ ,\\
s\lb &=&c.c. \ ,\\
sF&=&\fr [\c\6 F+\cb\deb F]+\2 \left[\6 c-\frac{\cb}{\1 }\6 \mu\right] F
            \non \\
 & &  +\2 \left[\deb\bar{c}-\frac{\c}{\1 }\deb\mb\right] F  \non \\
 & & - \frac{i}{2}\left[\left(\frac{\a\cb}{\1 }+\e\right) D_z\lb -\left(
    \f{\ab\c}{\1 }+\eb\right) D_{\zb}\l\right] \ ,  \\
s\mu &=&(\deb -\mu\6 +\6 \mu )c+\2 \a\e \ ,\\
s\mb &=&(\6 -\mb\deb +\deb\mb )\bar{c}+\2 \ab\eb \ ,\\
s\a &=&(\deb -\mu\6 +\2 \6 \mu )\e +c\6 \a +\2 \a\6 c \ ,\\
s\ab &=&(\6 -\mb\deb +\2 \deb\mb )\eb +\bar{c}\deb\ab
              +\2 \ab\deb\bar{c} \ ,
\eeqa
and
\beqa
sc&=&c\6 c-\frac{1}{4}\e\e \ ,\\
s\bar{c}&=&\bar{c}\deb \bar{c}-\frac{1}{4}\eb\eb \ ,\\
s\e &=&c\6 \e - \2 \e\6 c \ ,\\
s\eb &=&\bar{c}\deb\eb - \2 \eb\deb\bar{c} \ ,
\eeqa
so that
\beq
s^2 =0 \ ,
\eeq
with the supercovariant derivatives $D_z X$ and $D_z \lb$ defined as
\cite{deld}
\beqa
D_z X&\equiv&\fr [\de X+\2 \mb\a\l -\2 \ab\lb ] \ ,\\
D_z \lb&\equiv&\fr [(\6 -\mb\deb -\2 \deb\mb )\lb
      -\2 \ab (D_{\zb}X)+\frac{i}{2}\mb \a F] \ .
\eeqa
The fields $(c,\bar{c})$ have been introduced by Becchi \cite{becc} and
are related to the usual diffeomorphism ghosts $(\xi ,\bar{\xi})$ by
\beqa
c=\xi +\mu\bar{\xi} \ ,\non\\
\bar{c}=\bar{\xi}+\mb\xi \ .
\eeqa
Their superpartners $(\e,\eb)$ have bosonic statistic and are
related in a similar way
to the superdiffeomorphism ghosts $(\xi^{\t},\bar{\xi^{\t}})$ by
\cite{deld}
\beqa
\e=\xi^{\t}+\bar{\xi}\a \ ,\non\\
\eb =\bar{\xi^{\t}}+\xi\ab \ .
\eeqa
In order to fix the superdiffeomorphism invariance of (\ref{invaction})
we use a superconformal gauge. For the corresponding gauge fixing
term one has
\beq
-2iS_{gf}=\in \Lp bs\mu +\bar{b}s\mb -b^{\a}s\a -\bar{b}^{\a}s\ab \Rp \ ,
\eeq
where $(b,\bb)$ are the usual antighost fields, whereas $(\balf ,\bab)$
are the corresponding supersymmetric partners possessing now a bosonic
statistics and
\beq
  s b= s\bb = s\balf = s\bab =0 \ .
\eeq
To translate the BRS invariance of the gauge fixed action
$(S_{inv} + S_{gf})$ into a Slavnov identity we introduce
a set of invariant external sources $(\X ,\L ,\Lb ,\F ,\C
,\Cb ,\E ,\Eb)$ coupled to the nonlinear BRS variations of the
fields, i.e.
\beq
-2iS_{ext}=\in \Lp \X sX+\L s\l +\Lb s\lb +\F sF+\C sc
             +\Cb s\bar{c}+\E s\e +\Eb s\eb \Rp \ .
\eeq
The complete action
\beq
\S =S_{inv}+S_{gf}+S_{ext} \ ,
\eeq
obeys then to the Slavnov identity
\beq
  S(\S )  = 0  \ ,
\label{slavnov}
\eeq
with
\beqa
S(\S )=\in \Lp \f{\ds}{\d\X}\f{\ds}{\d X}+ \f{\ds}{\d\L}\f{\ds}{\d\l}+
 \f{\ds}{\d\Lb}\f{\ds}{\d\lb}+ \f{\ds}{\d\F}\f{\ds}{\d F}+
 \f{\ds}{\d\C}\f{\ds}{\d c}+ \f{\ds}{\d\Cb}\f{\ds}{\d\bar{c}} \non\\
+\f{\ds}{\d\E}\f{\ds}{\d\e}+ \f{\ds}{\d\Eb}\f{\ds}{\d\eb}+
 \f{\ds}{\d b}\f{\ds}{\d\mu}+ \f{\ds}{\d\bb}\f{\ds}{\d\mb}+
 \f{\ds}{\d\balf}\f{\ds}{\d\a} +
  \f{\ds}{\d\bab}\f{\ds}{\d\ab} \Rp  \ .
\eeqa
{}From the above identity one can read off the linearized Slavnov
operator ${\cal B}$
\beqa
{\cal B}=\in\Lp  \f{\ds}{\d\X}\f{\d}{\d X}+\f{\ds}{\d X}\f{\d}{\d\X}+
\f{\ds}{\d\L}\f{\d}{\d\l}+\f{\ds}{\d\l}\f{\d}{\d\L}+
\f{\ds}{\d\Lb}\f{\d}{\d\lb}+\f{\ds}{\d\lb}\f{\d}{\d\Lb} \non\\
+\f{\ds}{\d\F}\f{\d}{\d F}+\f{\ds}{\d F}\f{\d}{\d\F}+
\f{\ds}{\d\C}\f{\d}{\d c}+\f{\ds}{\d c}\f{\d}{\d\C}+
\f{\ds}{\d\Cb}\f{\d}{\d\bar{c}}+\f{\ds}{\d\bar{c}}\f{\d}{\d\Cb} \non\\
+\f{\ds}{\d\E}\f{\d}{\d\e}+\f{\ds}{\d\e}\f{\d}{\d\E}+
\f{\ds}{\d\Eb}\f{\d}{\d\eb}+\f{\ds}{\d\eb}\f{\d}{\d\Eb}+
\f{\ds}{\d b}\f{\d}{\d\mu}+\f{\ds}{\d\mu}\f{\d}{\d b} {\ }{\ }
{\ }{\ }{\ }{\ } \non\\
+\f{\ds}{\d\bb}\f{\d}{\d\mb}+\f{\ds}{\d\mb}\f{\d}{\d\bb}+
\f{\ds}{\d\balf}\f{\d}{\d\a}+\f{\ds}{\d\a}\f{\d}{\d\balf}+
\f{\ds}{\d\bab}\f{\d}{\d\ab}+\f{\ds}{\d\ab}\f{\d}{\d\bab} \Rp \ ,
\eeqa
which, as a consequence of the Slavnov identity, turns out to be
nilpotent
\beq
{\cal B \cal B}=0 \ .
\eeq
As explained in \cite{baul,deld} the Beltrami(no) parameters are treated
as unquantized external fields.
In particular $(\mu,\mb)$ are the sources for the components
$(T_{zz},T_{\zb\zb})$ of the energy-momentum tensor, i.e.
\beq
T_{zz}      =\f{\ds}{\d\mu}\quad ,\quad
T_{\zb\zb}  =\f{\ds}{\d\mb}  \ .
\label{energy}\eeq
In the same way the two Beltramino fields $(\a, \ab)$ describe the
fermionic counterpart of (\ref{energy}):
\beq
T_{\a}     =\f{\ds}{\d\a}\quad ,\quad
T_{\ab}  =\f{\ds}{\d\ab}\quad .
\eeq
One sees then that the classical Slavnov identity (\ref{slavnov})
describes the current algebra of the components of
the energy-momentum tensor.
In particular, from the expression for the linearized Slavnov operator
$\cal B$ it follows
\beq
T_{zz}  ={\cal B}b\quad ,\quad T_{\zb\zb} ={\cal B}\bb\quad ,\quad
T_{\a} ={\cal B}\balf\quad ,\quad T_{\ab} ={\cal B}\bab\quad \ .
\label{2}\eeq
meaning that the energy momentum tensor is cohomologically trivial.
In addition, it is easily verified that the complete action $\S$ is
itself a $\cal B$-variation
\beq
-2i\S ={\cal B}\in \Lp \2 \X X-\2 \L\l -\2 \Lb\lb
+\2 \F F-\C c-\Cb\bar{c}+
\E\e +\Eb\eb \Rp \ .\label{3}
\eeq
Equations (\ref{2}) and (\ref{3}) are the generalization to superstring
of the algebraic properties already found in the bosonic string
\cite{sore}. They suggest that, as in the bosonic case, one can interpret
the superstring as a topological model of the Witten's type \cite{witt}.

\section{The vector supersymmetry}
\setcounter{equation}{0}
In order to show the existence of the vector supersymmetry we introduce
the two functional operators $W$ and $\bar{W}$
\beqa
W=\in \Lp \mb\f{\d}{\d\bar{c}}+\f{\d}{\d c}+\Cb\f{\d}{\d\bb}+\ab\f{\d}{\d\eb}
+\Eb\f{\d}{\d\bab} \Rp \ ,\\
\Wb =\in \Lp \mu\f{\d}{\d c}+\f{\d}{\d\bar{c}}+\C\f{\d}{\d b}+\a\f{\d}{\d\e}
+\E\f{\d}{\d\balf} \Rp \ .
\eeqa
These operators are a direct extension of the vector supersymmetry
operators introduced for the bosonic string \cite{sore} and, together
with the linearized operator $\cal B$, give rise to the following
algebraic relations
\beq
\{\B ,W\} =\6 \quad,\quad\{\B ,\Wb\} =\deb\quad,\quad\{ W ,W\} =\{ W,\Wb\} =
\{\Wb ,\Wb\} =0 \ .\label{4}
\eeq
The algebra (\ref{4}) closes on the space-time translations, it this then
a supersymmetric algebra of the Wess-Zumino type. Eqs.(\ref{4})
represent a typical feature of the topolgical models \cite{guad,lucc} and,
as we will see in the next section, give a simple method for solving the
descent equations corresponding to the superdiffeomorphism anomaly.

Let us observe also that, in complete analogy with the bosonic case,
the operators $W$ and $\Wb$ when applied to the classical action $\S$
yield the linearly broken Ward identities
\beq
W\S =\D \quad,\qquad\Wb\S =\Db \ ,
\eeq
with
\beqa
\D =\f{i}{2}\in\Lp \C\6 c+\Cb\6 \bar{c}-\E\6 \e
       -\Eb\6 \eb -\X\6 X+\L\6 \l
\non\\
  +\Lb\6 \lb -\F\6 F-b\6 \mu
   -\bb\6 \mb +\balf\6 \a +\bab\6 \ab \Rp \ ,\\
\Db =\f{i}{2}\in\Lp \C\deb c+\Cb\deb\bar{c}-\E\deb\e
         -\Eb\deb\eb -\X\deb X+\L\deb\l \non\\
+\Lb\deb\lb -\F\deb F-b\deb\mu
-\bb\deb\mb +\balf\deb\a +\bab\deb\ab \Rp \ .
\eeqa
Notice that $\D$ and $\Db$, being linear in the quantum
fields, are classical breakings \cite{pigu}.

\section{The superdiffeomorphism anomaly}
\setcounter{equation}{0}
In this section we apply the supersymmetric structure (\ref{4}) in order
to solve the BRS consistency conditions for the Slavnov anomaly.
In the following we identify the string world sheet with the
complex plane $C$, the result can be generalized to any Riemann surface by
introducing an appropriate projective connection, as done
in \cite{stor,lazz}.

At the quantum level the classical action $\S$ gives rise to a
one-loop effective action
\beq
\Gamma =\S +\hbar\Gamma^{(1)} \ ,
\eeq
which obeys the anomalous Slavnov identity
\beq
S(\Gamma )=\hbar \AA \ .
\eeq
The anomaly $\AA$ is an integrated local two form with ghost number one
\footnote{ We adopt here the usual convention of
denoting with $\AA_{q}^{p}$ a $q$-form with ghost number equal to $p$. }
\beq
 \AA = \int \AA^1_2
\eeq
which has to fulfill the BRS consistency condition
\beq
 {\cal B} \AA = 0 \ .
\eeq
This condition, when translated to the nonintegrated level, yields a
tower of descent equations
\beqa
\B \AA^1_2  + d \AA^2_1 =0 \ ,\non\\
\B \AA^2_1  + d \AA^3_0 =0 \ ,\label{ladder}\\
\B \AA^3_0 =0 \ ,\non
\eeqa
where $d$ denotes the exterior space-time differential
\beq
d=d_z \6 +d_{\zb}\deb\quad ,\quad d^2 =0\quad ,\quad\{ \B ,d\} =\{ W,d\} =
\{ \Wb ,d\} =0 \ .
\eeq
As shown in \cite{sore}, in order to find a solution of the ladder
(\ref{ladder}) it is sufficient to know the nontrivial solution of the
last equation of the tower (\ref{ladder}). It is easy to check indeed
that, once a nontrivial solution for $\AA^3_0$ has been found, the
cocycles $\AA^2_1$ and $\AA^1_2$ are obtained by successive applications
of the operators $W$ and $\Wb$ on $\AA^3_0$, i.e.
\beqa
\AA^2_1 =(W \AA^3_0 )dz+(\Wb \AA^3_0 )d\zb \ ,\\
\AA^1_2 = (\Wb W \AA^3_0 )dz\wedge d\zb \ .
\label{5}
\eeqa
For what concerns the local cohomology of the linearized operator
$\cal B$ in the zero form sector with ghost number three it turns out
that the relevant cocycle $\AA^3_0$ can be identified, modulo a
$\cal B$ coboundary, with
\beq
 \AA^3_0 =c\6 c\6 ^2 c-c\6 \e\6 \e +\2 \6 c\6 \e\e -\bar{c}\deb\bar{c}
\deb ^2\bar{c}+\bar{c}\deb\eb\deb\eb -\2 \deb\bar{c}\deb\eb\eb \ .
\eeq
This expression is the supersymmetric extension of the well known
term $\AA_{0(bos.)}^3$
\beq
\AA_{0(bos.)}^3 =   c\6 c\6 ^2 c -\bar{c}\deb\bar{c}\deb ^2\bar{c} \ ,
\eeq
which is at the origin of the diffeomorphism anomaly for the bosonic
string \cite{baul,sore,lazz}.

Applying now the formula (\ref{5}) to $\AA^3_0$, for the
integrated Slavnov anomaly $\AA$ one gets the expression
\beq
\AA = \in \AA^1_2 = 2\in (c\6 ^3 \mu +\e\6 ^2 \a )
                    +2\in (\bar{c}\deb ^3 \mb +\eb\deb ^2\ab ) \ .
\label{anom11}
\eeq
This is the superdiffeomorphism anomaly (in the Wess-Zumino gauge)
of the (1,1) superstring theory \cite{deld}.
Using the truncation procedure of Sect. 2, for the $(1,0)$ case
one has
\beq
\AA_{(1,0)}= 2\in (c\6 ^3 \mu +\e\6 ^2 \a )
         +2\in (\bar{c}\deb ^3 \mb ) \ .
\label{anom10}
\eeq
Expressions (\ref{anom11}) fixes, through the numerical coefficient
of the corresponding Feynman diagrams,
the critical dimension of the $(1,1)$ superstring to be equal to 10, while
in the $(1,0)$ case Eq. (\ref{anom10}) leads to
a critical dimension $10$ in the supersymmetric sector and $26$ in the
nonsupersymmetric one.


\begin{thebibliography}{99}

\bibitem{witt} E. Witten, {\it Comm. Math. Phys.} {\bf 117} (1988) 353;\\
                          {\it Comm. Math. Phys.} {\bf 121} (1989) 351;
\bibitem{schw} A. S. Schwarz, {\it Lett. Math. Phys.} {\bf 2} (1978) 247;\\
     A. S. Schwarz, Baku International Topological Conference, Abstract,
     vol. 2, p. 345 (87);
\bibitem{birm} D. Birmingham, M. Blau, M. Rakowski, G. Thompson, {\it
     Phys. Rep.} {\bf 209} (1991) 129;
\bibitem{gier} F. Delduc, F. Gieres, S. P. Sorella, {\it Phys. Lett.}
     {\bf B225} (1989) 367;\\
     F. Delduc, C. Lucchesi, O. Piguet, S. P. Sorella, {\it Nucl. Phys.}
     {\bf B346} (1990) 313;
\bibitem{guad} E. Guadagnini, N. Maggiore, S. P. Sorella, {\it Phys. Lett.}
     {\bf B255} (1991) 65;\\
     S. P. Sorella, {\it Comm. Math. Phys.}  {\bf 157} (1993) 231;
\bibitem{lucc} C. Lucchesi, O. Piguet, S. P. Sorella,
     {\it Nucl. Phys.} {\bf B395} (1993) 325;
\bibitem{rako} D. Birmingham, M. Rakowski, {\it Phys. Lett.} {\bf B269}
     (1991) 103;\\
     {\it Phys. Lett.} {\bf B272} (1991) 217;\\
     {\it Phys. Lett.} {\bf B275} (1992) 289;\\
     {\it Phys. Lett.} {\bf B289} (1992) 271;\\
\bibitem{sore} G. Bandelloni and S. Lazzarini,
  {\it Diffeomorphism Cohomology in Beltrami parametrization},
 PAR-LPTM-93, GEF-TH/93; \\
 M. Werneck de Oliveira, M. Schweda and S. P. Sorella,
     {\it Phys. Lett.} {\bf B315} (1993) 93;
\bibitem{bbg} L. Baulieu, M. Bellon and R. Grimm,
                {\it Phys.Lett.} {\bf B198} (1987) 343; \\
                {\it Nucl. Phys.} {\bf B321} (1989) 697;
\bibitem{deld} F. Delduc, F. Gieres, {\it Class. Quant. Grav.} {\bf 7}
     (1990) 1907;
\bibitem{baul} L. Baulieu, C. Becchi, R. Stora, {\it Phys. Lett.} {\bf B180}
     (1986) 55;
     L. Baulieu, M. Bellon, {\it Phys. Lett.} {\bf B196} (1987) 142;
\bibitem{becc} C. Becchi, {\it Nucl. Phys.} {\bf B304} (1988) 513;
\bibitem{bela} A. A. Belavin, A. M. Polyakov, A. B. Zamolodchikov,
     {\it Nucl. Phys.} {\bf B241} (1984) 333;
\bibitem{bagg} J. Wess, J. Bagger, {\it Supersymmetry and Supergravity},
     Princeton University Press, Princeton, NJ (1983);
\bibitem{pigu} O. Piguet, S. P. Sorella, {\it Helv. Phys. Acta} {\bf 63}
     (1990) 683;
\bibitem{stor} R. Stora, in {\it Non Perturbative Quantum Field Theory,} G.
     't Hooft and al. eds., Nato ASI series {\bf B}, vol. {\bf 185}, Plenum
     Press (88);\\
     S. Lazzarini and R. Stora, {\it Ward Identities for Lagrangian Conformal
     Models}, in {\it Knots, Topology and Quantum Field Theory}, $13^{th}$
     John Hopkins Workshop, L. Lusanna ed., World Scientific (89);
\bibitem{lazz} S. Lazzarini, {\it On Bidimensional Lagrangian Conformal
     Models}, Thesis LAPP-Annecy, France, (90) unpublished; \\
    R.Zucchini, {\it Comm. Math. Phys.} {\bf 152} (1993) 269;


\end{thebibliography}
\end{document}